**Saggu, A. (2022). The Intraday Bitcoin Response to Tether Minting and Burning Events: Asymmetry, Investor Sentiment, And "Whale Alerts" On Twitter. Finance Research Letters, 49, 103096. https://doi.org/10.1016/j.frl.2022.103096**

*Following publication, this research garnered significant attention and remains featured on Whale Alert's website under the title "The Whale Alert Effect." The article was prominently pinned to the top of Whale Alert's Twitter page for several months. The feature also included a guest research article I authored, which was kindly hosted by the Blockchain Research Lab in Germany. It is important to note that I had no prior association with Tether, Whale Alert, or the Blockchain Research Lab before publishing this paper.*

# The Intraday Bitcoin Response to Tether USD₮ Minting and Burning Events: Asymmetries, Investor Sentiment and Whale Alert Announcements on Twitter

Aman Saggu*

*Business Administration Division, Mahidol University International College*

**Abstract**

Tether Limited has the sole authority to create (mint) and destroy (burn) Tether stablecoins (USD₮). This paper investigates Bitcoin's response to USD₮ supply change events between 2014 and 2021 and identifies an interesting asymmetry between Bitcoin's responses to USD₮ minting and burning events. Bitcoin responds positively to USD₮ minting events over 5- to 30-minute event windows, but this response begins declining after 60 minutes. State-dependence is also demonstrated, with Bitcoin prices exhibiting a greater increase when the corresponding USD₮ minting event coincides with positive investor sentiment and is announced to the public by data service provider, Whale Alert, on Twitter.

*Corresponding author: Aman Saggu, Business Administration Division, Mahidol University International College, 999 Phutthamonthon Sai 4 Rd, Salaya, Phutthamonthon District, Nakhon Pathom 73170, Thailand, Telephone: 0066 27005000, Fax: 0066 24415091, Email address: aman.sag@mahidol.edu; asaggu26@gmail.com

> *"Questions about the influence of Tether continue to swirl in cryptocurrency markets".*
>
> *~Bloomberg* (Ossinger, 2021)

**1. Introduction**

Decentralized finance (DeFi) has revolutionized the cryptocurrency economy, enabling peer-to-peer financial services without the participation of financial intermediaries (Brooks, 2021). Despite these innovations, cryptocurrency markets continue to experience considerable volatility compared to traditional financial markets (Borgards and Czudaj, 2020). This has given rise to the stablecoin—a class of cryptocurrency and major application of DeFi. Issuers create stablecoins on blockchains and peg valuations to more stable external references such as a fiat currency, commodity, or basket of assets. They provide lower transaction costs and greater transaction speeds than traditional banks and payment providers (BIS, 2019). The asset-backed stablecoin sub-class derives value from trust that they are fully collateralized by reserves of assets held by issuers. The seignorage-style stablecoin sub-class derives value from user demand, and maintains the peg using algorithms governing the stablecoin's supply.[1] Although stablecoins are used almost exclusively for cryptocurrency-related transactions, they face growing institutional adoption. J.P. Morgan tried their "JPM stablecoin" in October 2020; Visa added stablecoins to their settlement platform in March 2021; and the British government outlined plans to recognize stablecoins as a valid form of payment in April 2022.

---

[1] It is beyond the scope of this study to validate proofs of reserves which are sometimes in question. The 1:1 redemption policies differ by stablecoin and may evolve over time. Tether Limited initially claimed that minted USD₮ were backed by USD held in reserve, and 1:1 redemptions for USD were not guaranteed. Following a legal case, Tether Limited revised claims to state that minted USD₮ were backed and collateralized by a complex reserve of assets, and not fiat USD alone. Tether USD₮ can today be redeemed for USD at a 1:1 ratio from Tether Limited. Arbitrageurs can use this mechanism to purchase USD₮ trading under par, and redeem them at a profit, thereby supporting the peg.

Tether is the most widely traded stablecoin in the world.[2] By December 2021, Tether had a market capitalization exceeding $70 billion United States dollars (USD), placing it among the largest cryptocurrencies globally by market capitalization. Tether was created in 2014 by a centralized entity, Tether Limited, which has the sole authority to mint (create) and burn (destroy) Tether stablecoins—denoted as "USD₮"—which enables the company to add and remove liquidity from the cryptocurrency ecosystem.[3] On the supply side, Tether Limited routinely mint (burn) USD₮ when demand for the stablecoin exceeds (subceeds) supply in the open market (Floyd, 2018). The demand for USD₮ principally stems from investors, as it is the preeminent stablecoin used to buy and sell Bitcoin and digital assets on cryptocurrency exchanges. USD₮ is also widely used to transact in the DeFi ecosystem. We therefore hypothesize that news of USD₮ minting (burning) events are perceived by market participants as positive (negative) signals concerning underlying demand for cryptocurrency assets such as Bitcoin and DeFi services. We formulate and test three hypotheses: (H1): USD₮ minting has a positive impact on Bitcoin returns; (H2): USD₮ burning has a negative impact on Bitcoin returns; (H3): Effects (H1) and (H2) are similar in magnitude.

Tether Limited can mint and burn USD₮ on various blockchains at any time without prior notice. These irregular events are well-suited for event-study analyses that compare ex-ante and ex-post market responses.[4] Wei (2018) found no evidence that USD₮ minting influenced Bitcoin returns. Ante et al. (2021) revealed mixed evidence of significantly abnormal Bitcoin returns occurring during the 24 hours immediately before a USD₮ minting event. No significantly abnormal returns were observed during the 12 hours after a minting event, but weakly significant abnormal returns (0.63%) were noted 24 hours after. Griffin and Shams (2020) observed that Bitcoin only responded positively to USD₮ minting events following negative price shocks, which they attributed to a market manipulator. They also found that months with larger USD₮ minting events were followed by negative Bitcoin returns, which was interpreted as Tether Limited liquidating Bitcoin holdings to satisfy end-of-month reporting requirements. These findings were contested by Lyons and Viswanth-Natraj (2020), who found no systematic evidence that USD₮ minting events impacted Bitcoin prices.

---

[2] See https://www.kraken.com/en-gb/learn/what-is-tether-usdt.

[3] Tether was formerly known as Realcoin and was later re-branded as Tether.

[4] Terminology pertaining to the creation of new USD₮ is blockchain-specific. However, each minted USD₮ represents $1 USD and is in principle equivalent across blockchains, although not transferrable between blockchains. Each USD₮ which is burned, revoked or destroyed and destroyed is effectively removed from circulation and/or declared non-redeemable and flagged. Although these abilities are technically different, they serve the same objective as to be made non-usable.

To advance upon previous studies using intraday data, we identify Bitcoin's response to every USD₮ minting event and USD₮ burning event on all blockchains used by Tether Limited that occurred between 2014 and 2021. This is the largest dataset of USD₮ events to be used in research. Our results can be summarized as follows. Firstly, Bitcoin responds significantly to USD₮ minting events, particularly, over 5- to 30-minute event windows. Secondly, after accounting for conditionality in event type, we reveal asymmetries: there is a significant (insignificant) response to USD₮ minting (burning) events. This is consistent with the fear-of-missing-out (FOMO) phenomenon (Baur and Dimpfl, 2018), as Bitcoin responds to positive news (USD₮ minting) but not to negative news (USD₮ burning) events. Thirdly, motivated by previous studies that have highlighted the importance of investor sentiment in the Bitcoin pricing paradigm (Bouri et al., 2021; Anamika et al., 2021), we demonstrate that Bitcoin responds primarily to USD₮ minting events during periods of positive sentiment. Finally, we show that Bitcoin responds principally to USD₮ minting events when they occur simultaneously with positive investor sentiment and are announced to the public in a "Whale Alert" published by the eponymous cryptocurrency data service provider on Twitter. This study complements existing research regarding the role of influential Twitter accounts in the cryptocurrency space (Ozturk and Bilgic, 2021; Shahzad et al., 2022).

## 2. Data Background

*2.1 Tether (USD₮) Data*

The terminology for creating new USD₮ and removing existing USD₮ from circulation differs according to the blockchain. We refer to these activities as "minting" and "burning" events, respectively.[5] Our dataset spans from the very first USD₮ minting event, which occurred between October 6, 2014 18:54:05 and January 9, 2021 13:20:09 UTC. Data-concerning events are obtained by comprehensively inspecting Tether Treasury wallet transactions on the respective block explorer websites. We sourced Omni USD₮ data from omniexplorer.info, Ethereum USD₮ data from etherscan.io, Tron USD₮ data from tronscan.org, EOS USD₮ data from bloks.io, Liquid USD₮ data from blockstream.info, SLP USD₮ data from simpleledger.info, and Algorand USD₮ data from goalseeker.purestake.io. We validated the dataset by comparing the data to the cumulative total USD₮ balances published by Tether Limited.[6] Over the sample period, the greatest minting and burning events occurred symmetrically at 1 billion USD₮ and –1 billion USD₮, respectively.[7][8][9]

---

[5] Minting includes minting, granting and issuance. Burning includes burning, revocation and destruction.

[6] See https://wallet.tether.to/transparency.

[7] The dataset, as described in Table 1, merges 32 USD₮ minting and burning event-dates with identical timestamps on the same blockchains: 8 minting events combined to 4 minting events, and 24 burning events combined to 10 burning events. On 26 May 2017 21:48:40 UTC, Tether Limited minted 10 million Omni USD₮, and then minted another 10 million Omni USD₮. We combine these two events into a minting of 20 million Omni USD₮ because they had identical timestamps on the same blockchain. We identify a total of 32 event-dates with simultaneous minting or burning, with identical timestamps on the same blockchains. In some cases, more than two burning events occurred at the same time on the same blockchain. For example, on 08 Jul 2020 18:32:07 UTC, a total of 0.14 SLP USD₮ was created in four separate events. These are combined.

*2.2 Bitcoin Price (USD) Data*

Bitcoin price returns, $r_t$, are defined as the first difference of the natural log of the Bitcoin price, $Ƀ_t$, $n$ minutes after each USD₮ minting and burning event date, relative to 1 minute before each event, for 5-, 10-, 15-, 30-, 60- and 1,440-minute event windows. Returns are calculated using the following: $r_t = 100 * \ln(Ƀ_n/Ƀ_{n-1})$. Bitcoin prices at the 1-minute frequency are sourced from bitcoincharts.com. We use Bitfinex price data from the beginning of the sample period through December 22, 2016 and Bitstamp price data thereafter. This ensures the creation of a complete dataset that does not overlook any data points and facilitates its replicability using publicly available data.[10] The mean value of Bitcoin returns occurring around USD₮ minting events is positive and increases in magnitude from 5- to 30-minute windows (see Table 1). The mean value of Bitcoin returns occurring around USD₮ burning events is negative and increases in magnitude from 5- to 15-minute windows. This indicates that USD₮ minting (burning) events are, on average, followed by positive (negative) returns.

---

[8] We exclude the accidental minting and subsequent burning of 5 billion Tron USD₮: one minting and two burning event-dates – in line with Ante et al (2021). Tether Limited accidentally minted 5 billion Tron USD₮ on 13 Jul 2019 17:34:24 UTC. The error was quickly realised, and the 5 billion Tron USD₮ were subsequently burned within 15 minutes, thereby increasing the difficulty in isolating effects on the market. The sheer size of these of these observations exerts significant influence on estimates and inflates standard errors.

[9] We also exclude the simultaneous minting and burning of 1,008.97 SLP USD₮ on 03 Jun 2020. On 03 Jun 2020 13:23:53 UTC, Tether Limited simultaneously minted 1,008.97 SLP USD₮ and burned 1,008.97 SLP USD₮ at exactly the same time. We exclude this event-date as the sum of 0 SLP USD₮ was created in the simultaneous minting and burning. These were amongst the first USD₮ minting and burning events on SLP, and the small redeemable value of $1,008.97 USD suggests they were simply conducted for testing.

[10] Bitfinex data are publicly available until December 22, 2016. Bitcoin was thinly (heavily) traded on Bitstamp (Bitfinex) at the beginning of the sample period.

## 3. Econometric Models and Results

3.1 *Baseline Analysis*

We begin by regressing Bitcoin returns on event dates when there is a change in the USD₮ supply. A change in USD₮ supply, $\Delta T_t$, is measured in billions of USD₮.[11] $\alpha$ is the intercept of the regression. $e_t$ is the error term. Panel (a) of Table 2 presents the ordinary least squares (OLS) estimates of Equation (1) with Newey–West robust standard errors.

$$r_t = \alpha + \beta_1 \Delta T_t + e_t \qquad (1)$$

Bitcoin's response to USD₮ supply changes is statistically significant for 5- to 60-minute event windows. A positive value for the $\beta_1$ coefficient indicates that a USD₮ minting (burning) event was perceived positively (negatively) by short-term Bitcoin investors. The minting (burning) of 1 billion (–1 billion) USD₮ results in a 0.24% increase (decrease) in Bitcoin price in a 5-minute window. This rises to a 0.38% increase (decrease) in a 10-minute window, 0.51% increase (decrease) in a 15-minute window, and peaks at a 0.68% increase (decrease) in a 30-minute window before declining to a 0.57% increase (decrease) in a 60-minute window. As a robustness check, the maximum likelihood (MM)-weighted least squares estimates are presented in Panel (b) of Table 2.[12][13]

These results align with our hypotheses (H1) and (H2): news concerning a USD₮ minting (burning) event has a positive (negative) impact on Bitcoin returns. Previous event-studies have observed little to no significant Bitcoin response to USD₮ minting events using hourly or daily data (Wei, 2018; Ante et al., 2021; Lyons and Viswanth-Natraj, 2020). We encounter a weak or insignificant response using 60- and 1,440-minute windows. We contribute to literature by demonstrating that the largest impacts occur during short-term (5- to 30-minute) event windows.[14] That the strong, instantaneous price impact of news dissipates within 60 minutes is consistent with observations of the gold market (Elder et al., 2012).[15] The rapid market response may also indicate high-speed trading.

---

[11] An increase (decrease) in the supply of USD₮ is considered a minting (burning) event.

[12] MM estimates are more robust in the presence of large outliers. The adjusted rw-squared statistic of Renaud and Victoria-Feser (2010) is reported for MM estimates.

[13] Following the peer reviewers' suggestion, we controlled for changes in the volatility index (VIX), the price of gold and the Standard & Poor's 500 Index in line with Corbet et al. (2020). We further controlled for the Hedonometer Happiness Index (Bouri et al., 2021) and the Economic Policy Uncertainty/Economic Uncertainty-Related Queries Index (Bouri and Gupta, 2021). The estimates are highly robust to various model specifications using these variables. The results are available upon request.

*3.2 Asymmetries*

To account for the possibility that investors respond differently to USD₮ minting and burning events, we augment the model with two dummy variables, $D_t^m$ and $D_t^b$, which equal one when the USD₮ supply change is due to a minting or burning event and zero otherwise.

$$r_t = \alpha + (\beta_1 D_t^m + \beta_2 D_t^b)\Delta T_t + e_t \qquad (2)$$

The OLS and MM-estimates of Equation (2) presented in Table 3 reveal a notable asymmetry after accounting for conditionality in event type. The positivity of the $\beta_1$ coefficient indicates that Bitcoin investors respond positively and significantly to USD₮ minting events. The response magnitude increases between the 5- and 30-minute windows. In contrast, the $\beta_2$ coefficient is mostly positive but insignificant for all event windows. This indicates that the burning of −1 billion USD₮ has a negative but insignificant impact on Bitcoin returns. Wald tests results $H_0: (\beta_1 = \beta_2)$ support the fact that there are significantly divergent responses to USD₮ minting and USD₮ burning events. All subsequent tables report a comprehensive series of Wald tests for each model.

---

[14] Event studies analyzing asset responses to news announcements typically employ tighter intraday event windows to isolate the effects from other potentially market-moving news (Bianchi et al., 2021).

[15] Bouri et al. (2019b) found that price explosiveness in Bitcoin persists for a longer duration than it does in other cryptocurrencies.

That USD₮ minting (burning) events were interpreted as positive (non-response) by Bitcoin investors is consistent with the FOMO phenomenon—an asymmetric, irrational psychological behavior observed in speculative investors documented in the broader cryptocurrency research. Güler (2021) demonstrated that positive news had a greater impact on Bitcoin returns than negative news before the emergence of COVID-19. Baur and Dimpfl (2018) similarly found that positive news shocks were associated with higher Bitcoin volatility than negative news shocks. Both studies concluded that these asymmetrical responses, which were biased toward positive news were consistent with the FOMO phenomenon being prevalent in the cryptocurrency space. This asymmetry—there is a positive response to a USD₮ minting event (positive news) but a non-response to a USD₮ burning event (negative news)—is consistent with the effects of the FOMO phenomenon (see Table 3). Bouri et al. (2017) proposed an alternative explanation that such asymmetries indicate Bitcoin's safe-haven properties.

Motivated by these findings, we continued our investigation by explicitly accounting for investor psychology. We augmented the model with slope interactive variables to capture state-dependence in the relationship between Bitcoin returns and USD₮ minting and burning events with respect to "positive" and "negative" investor sentiment measures as follows:

$$r_t = \alpha + [\beta_1(D_t^m D_t^+) + \beta_2(D_t^m D_t^-) + \beta_3(D_t^b D_t^+) + \beta_4(D_t^b D_t^-)]\Delta T_t + e_t \qquad (3)$$

where $D_t^+$ and $D_t^-$ in turn take the form of dummy variables equal one (zero) when (a) the Crypto Fear & Greed Index points to extreme greed or greed (extreme fear or fear); (b) the Sentix Bitcoin Institutional Investor Sentiment Indicator surveys more bulls than bears (more bears than bulls); and (c) the Lunde and Timmermann filter (2012) indicates a bullish (bearish) market.[16]

---

[16] The so-called "Crypto Fear & Greed Index" maintained by alternative.me measures the short-term daily emotional state of cryptocurrency investors. The factors determining its value include volatility, market

The estimates of Equation (3) presented in Table 4 show that Bitcoin's response to USD₮ minting events during periods of positive sentiment ($\beta_1$) is positive and significant for 5- to 30-minute event windows and greater in magnitude than during periods of negative sentiment ($\beta_2$) using any of the three sentiment measures. Bitcoin investors do not respond to USD₮ burning events—negative news—during periods of positive ($\beta_3$) or negative investor sentiment ($\beta_4$).[17] This supports Karalevicius et al. (2018): intraday Bitcoin prices overreact to news events in the direction of the sentiment, followed by a correction. We similarly discover that investors' response to positive news—USD₮ minting events—during periods of positive sentiment is maximized in a 30-minute window, followed by a correction that occurs within 60 minutes. The weaker response during periods of negative sentiment is in line with Bouri et al. (2021): cryptocurrencies may be used for hedging when investor sentiment declines.

*3.3 Whale Alerts*

For investors to capitalize on Bitcoin price increases following USD₮ minting events, they require technical expertise and a computing infrastructure that enables them to continually monitor blockchains for relevant news-events. Alternatively, they may obtain such information from Whale Alert, a third-party information provider that monitors millions

---

momentum/volume, social media metrics, surveys, Bitcoin dominance and Google trends. A measurement of extreme fear or fear (extreme greed or greed) is indicative of Bitcoin's price being greater than (less than) its intrinsic value. The Sentix Institutional Investor Sentiment Indicator (SNTMXBH1) published by sentix.de measures the 1-month-ahead sentiment of institutional Bitcoin investors (Anamika et al., 2021). It is constructed using a weekly survey of up to 5,000 participants and subtracting the percentage of bullish investors from the percentage of bearish investors. The Lunde and Timmermann (2012) model is used to identify bullish and bearish states using daily Bitcoin price data from 2011 through 2021. The underlying latent states are defined using logged price changes from previous local peaks or troughs.

[17] Corbet et al. (2020) found that the theft of $30 million USD₮ on November 21, 2017 had no statistically significant impact on Bitcoin price volatility despite widespread news coverage. Bouri et al. (2019a) determined that trading volumes granger-caused extremely positive or negative returns.

of daily cryptocurrency transactions and publishes notable events on Twitter in near real time. Whale Alert published 222 of 367 USD₮ minting events (60%) and 11 of 220 (5%) burning events in our sample on Twitter. Whale Alert has an influential presence in the cryptocurrency space with over 1.4 million followers. To account for the possibility that Bitcoin investors respond to Whale Alert tweets regarding USD₮ minting events, we augmented the model with a dummy variable, $D_t^{wa}$, that equals one when Ethereum and Tron USD₮ minting events were accompanied by a Whale Alert tweet and zero otherwise.[18] Its role in the function can be seen in the following equation:

$$r_t = \alpha + [(\beta_1 D_t^+ + \beta_2 D_t^-)D_t^{wa} + (\beta_3 D_t^+ + \beta_4 D_t^-)(1 - D_t^{wa})]\Delta T_t D_t^m + e_t \qquad (4)$$

The estimates of Equation (4) presented in Table 5 show that Bitcoin investors responded significantly to USD₮ minting events when they occurred during a period of positive sentiment and simultaneously tweeted by Whale Alert ($\beta_1$) for the 5- to 30-minute event windows. The response is greater in magnitude compared to its negative sentiment counterpart with Whale Alert tweets ($\beta_2$). Bitcoin investors usually do not respond to USD₮ minting events that are not tweeted by Whale Alert during periods of positive ($\beta_3$) or negative sentiment ($\beta_4$). These results highlight the critical role that Whale Alert plays in informing investors about USD₮ minting events. This complements Ozturk and Bilgic (2021)—the informational content of tweets from influential accounts is an important predictor of Bitcoin returns.[19]

---

[18] On three occasions, Whale Alert Tweeted the transfer of USD₮ instead of the actual minting events. These errors occurred in relation to minting events on 11 Jun 19 10:15:04; 16 Jun 19 13:21:19; and 21 Jun 19 17:57:27. We include these events because they were still announced as minting events on Twitter by Whale Alert.

[19] Shahzad et al. (2022) demonstrated that Elon Musk's tweets can be used by investors to predict periods of price explosivity.

## 4. Conclusion

This study date and time stamps every minting and burning event of Tether USD₮ stablecoins during the 2014–2021 period and reveals that Bitcoin principally responds to minting events of Tether USD₮ stablecoin when they (1) occur during a period of positive investor sentiment and (2) are announced to the public by a Whale Alert on Twitter. Bitcoin does not respond significantly to burning events coinciding with any other type of investor sentiment, even if they are tweeted by Whale Alert, which is consistent with the FOMO phenomenon. For investors, this indicates an opportunity to capitalize on short-term increases in the price of Bitcoin during the 5- to 30-minute windows following a USD₮ minting event, before the effects dissipate (within 60 minutes). Policymakers considering the adoption of a central bank digital currency should know that stablecoins originated in the cryptocurrency space and could therefore transmit irrational market behaviors, such as investor behavior characterized by the FOMO phenomenon, into traditional markets. Future studies should consider this impact on a broader range of cryptocurrencies, analyse feedback loops and causalities, consider the role of herding, and examine the impact on volatility connectedness.

# Tables and Figures

**Table 1: Descriptive statistics for Bitcoin returns occurring around USD₮ supply change events**

| Event Window | Obs. | Min | Max | Mean | Mode |
|---|---|---|---|---|---|
| *(a) Minting Events* | | | | | |
| 5-min | 367 | −1.724 | 2.131 | 0.072 | 0.000 |
| 10-min | 367 | −3.767 | 3.376 | 0.076 | 0.612 |
| 15-min | 367 | −2.115 | 5.109 | 0.104 | 0.000 |
| 30-min | 367 | −2.756 | 6.153 | 0.196 | 1.004 |
| 60-min | 367 | −6.642 | 7.441 | 0.154 | −0.081 |
| 1,440-min | 367 | −33.798 | 25.263 | 0.611 | 1.951 |
| *(b) Burning Events* | | | | | |
| 5-min | 220 | −0.453 | 0.987 | −0.003 | −0.026 |
| 10-min | 220 | −0.774 | 0.495 | −0.015 | −0.025 |
| 15-min | 220 | −0.773 | 1.366 | −0.021 | −0.008 |
| 30-min | 220 | −1.095 | 2.198 | 0.001 | −0.148 |
| 60-min | 220 | −1.458 | 4.056 | 0.001 | 0.026 |
| 1,440-min | 220 | −6.653 | 11.253 | −0.278 | 0.549 |

*Notes: Table 1 reports the descriptive statistics for Bitcoin returns around event dates on which there is a change in the USD₮ supply, with minting events reported in Panel (a) and burning events reported in Panel (b). The sample extends from October 6, 2014 18:54:05 UTC to January 9, 2021 13:20:09 UTC.*

**Table 2: Response of Bitcoin returns to USD₮ supply change events**

| Event Window | Obs. | α | $\beta_1$ | $\alpha^{SE}$ | $\beta_1^{SE}$ | Adj. $R^2$ |
|---|---|---|---|---|---|---|
| *(a) OLS Estimates* | | | | | | |
| 5-min | 587 | 0.03** | 0.24*** | (0.01) | (0.09) | 0.01 |
| 10-min | 587 | 0.03 | 0.38*** | (0.02) | (0.10) | 0.01 |
| 15-min | 587 | 0.04 | 0.51*** | (0.02) | (0.15) | 0.01 |
| 30-min | 587 | 0.09*** | 0.68*** | (0.03) | (0.21) | 0.01 |
| 60-min | 587 | 0.07* | 0.57* | (0.04) | (0.29) | 0.00 |
| 1,440-min | 587 | 0.23 | 1.19 | (0.19) | (0.92) | 0.00 |
| *(b) Robust MM-estimates* | | | | | | |
| 5-min | 587 | 0.01 | 0.29*** | (0.01) | (0.06) | 0.04 |
| 10-min | 587 | 0.02* | 0.41*** | (0.01) | (0.09) | 0.05 |
| 15-min | 587 | 0.02 | 0.44*** | (0.01) | (0.10) | 0.04 |
| 30-min | 587 | 0.04** | 0.57*** | (0.02) | (0.15) | 0.03 |
| 60-min | 587 | 0.03 | 0.37** | (0.02) | (0.16) | 0.01 |
| 1,440-min | 587 | 0.02 | 0.95 | (0.14) | (1.05) | 0.00 |

*Notes: Table 2 reports estimates of Equation (1) (i.e., $r_t = \alpha + \beta_1 \Delta T_t + e_t$) using ordinary least squares (OLS) regressions with Newey–West standard errors appearing in Panel (a) and those using maximum likelihood (MM)-estimates appearing in Panel (b). $r_t$ denotes Bitcoin returns; $\Delta T_t$ denotes the change in the USD₮ supply. The sample extends from October 6, 2014 18:54:05 UTC to January 9, 2021 13:20:09 UTC. Standard errors appear in parentheses. \*, \*\*, \*\*\* indicate significance at the 10%, 5% and 1% levels.*

**Table 3: Response of Bitcoin returns to USD₮ supply change events, controlling for minting and burning events**

| Event Window | Obs. | α | $\beta_1$ | $\beta_2$ | $\alpha^{SE}$ | $\beta_1^{SE}$ | $\beta_2^{SE}$ | $\beta_1=\beta_2$ | Adj. $R^2$ |
|---|---|---|---|---|---|---|---|---|---|
| *(a) OLS Estimates* | | | | | | | | | |
| 5-min | 587 | 0.03* | 0.33** | 0.06 | (0.01) | (0.14) | (0.05) | [0.09] | 0.01 |
| 10-min | 587 | 0.01 | 0.55*** | 0.05 | (0.02) | (0.14) | (0.08) | [0.00] | 0.01 |
| 15-min | 587 | 0.02 | 0.75*** | 0.06 | (0.02) | (0.21) | (0.08) | [0.00] | 0.02 |
| 30-min | 587 | 0.07** | 1.01*** | 0.05 | (0.03) | (0.32) | (0.10) | [0.01] | 0.02 |
| 60-min | 587 | 0.06 | 0.71* | 0.29 | (0.04) | (0.43) | (0.25) | [0.41] | 0.00 |
| 1,440-min | 587 | 0.20 | 1.66 | 0.31 | (0.21) | (1.42) | (0.79) | [0.43] | 0.00 |
| *(b) Robust MM-estimates* | | | | | | | | | |
| 5-min | 587 | 0.00 | 0.59*** | 0.02 | (0.01) | (0.08) | (0.11) | [0.00] | 0.09 |
| 10-min | 587 | 0.01 | 0.62*** | 0.04 | (0.01) | (0.11) | (0.16) | [0.00] | 0.07 |
| 15-min | 587 | 0.00 | 0.86*** | 0.03 | (0.01) | (0.12) | (0.17) | [0.00] | 0.09 |
| 30-min | 587 | 0.01 | 1.12*** | −0.05 | (0.02) | (0.18) | (0.26) | [0.00] | 0.07 |
| 60-min | 587 | 0.03 | 0.45** | 0.24 | (0.02) | (0.21) | (0.29) | [0.57] | 0.01 |
| 1,440-min | 587 | −0.01 | 1.42 | 0.22 | (0.15) | (1.32) | (1.86) | [0.61] | 0.00 |

*Notes: Table 3 reports estimates of Equation (2) (i.e., $r_t = \alpha + (\beta_1 D_t^m + \beta_2 D_t^b)\Delta T_t + e_t$) using OLS regressions with Newey–West standard errors appearing in Panel (a) and those using robust MM-estimates appearing in Panel (b). $r_t$ denotes Bitcoin returns; $\Delta T_t$ denotes the change in the USD₮ supply. $D_t^m$ and $D_t^b$ are dummy variables that equal one when the event date is a minting or a burning event and equal zero otherwise. The sample extends from October 6, 2014 18:54:05 UTC to January 9, 2021 13:20:09 UTC. Standard errors appear in parentheses. P-values from Wald tests (F-statistics) appear in square brackets. \*, \*\*, \*\*\* indicate significance at the 10%, 5% and 1% levels.*

**Table 4: Response of Bitcoin returns to USD₮ supply change events, controlling for minting and burning events and sentiment**

| Event Window | Obs. | α | β₁ | β₂ | β₃ | β₄ | α$^{SE}$ | β₁$^{SE}$ | β₂$^{SE}$ | β₃$^{SE}$ | β₄$^{SE}$ | [β₁=β₂] | [β₃=β₄] | [β₁=β₃] | [β₂=β₄] | Adj. R² |
|---|---|---|---|---|---|---|---|---|---|---|---|---|---|---|---|---|
| *(a) Crypto Fear and Greed Index of Alternative.me* | | | | | | | | | | | | | | | | |
| 5-min | 587 | −0.01 | 0.74*** | 0.17* | -0.04 | 0.01 | (0.01) | (0.09) | (0.09) | (0.16) | (0.16) | [0.00] | [0.82] | [0.00] | [0.38] | 0.14 |
| 10-min | 587 | 0.01 | 0.75*** | 0.14 | −0.06 | 0.06 | (0.01) | (0.12) | (0.13) | (0.22) | (0.21) | [0.00] | [0.70] | [0.00] | [0.74] | 0.10 |
| 15-min | 587 | 0.00 | 1.21*** | 0.15 | 1.39 | 0.03 | (0.01) | (0.13) | (0.14) | (0.93) | (0.24) | [0.00] | [0.16] | [0.85] | [0.66] | 0.15 |
| 30-min | 587 | 0.00 | 1.56*** | 0.08 | −0.15 | 0.09 | (0.02) | (0.20) | (0.21) | (0.37) | (0.36) | [0.00] | [0.64] | [0.00] | [0.98] | 0.13 |
| 60-min | 587 | 0.02 | 0.65*** | 0.15 | −0.19 | 0.39 | (0.02) | (0.23) | (0.24) | (0.43) | (0.41) | [0.11] | [0.33] | [0.09] | [0.62] | 0.03 |
| 1,440-min | 587 | −0.13 | 0.54 | 1.57 | 0.84 | −0.95 | (0.14) | (1.41) | (1.49) | (2.66) | (2.53) | [0.60] | [0.63] | [0.92] | [0.40] | 0.01 |
| *(b) Sentix Bitcoin Institutional Investor Sentiment Indicator* | | | | | | | | | | | | | | | | |
| 5-min | 587 | 0.00 | 0.79*** | 0.15 | −0.04 | 0.05 | (0.01) | (0.10) | (0.12) | (0.17) | (0.14) | [0.00] | [0.71] | [0.00] | [0.57] | 0.12 |
| 10-min | 587 | 0.01 | 0.70*** | 0.41** | −0.09 | 0.11 | (0.01) | (0.13) | (0.16) | (0.24) | (0.20) | [0.15] | [0.51] | [0.00] | [0.26] | 0.08 |
| 15-min | 587 | 0.01 | 1.01*** | 0.45** | −0.02 | 0.08 | (0.01) | (0.15) | (0.19) | (0.28) | (0.22) | [0.01] | [0.78] | [0.00] | [0.21] | 0.10 |
| 30-min | 587 | 0.02 | 1.23*** | 0.74*** | −0.17 | 0.03 | (0.02) | (0.22) | (0.27) | (0.40) | (0.32) | [0.14] | [0.70] | [0.00] | [0.10] | 0.07 |
| 60-min | 587 | 0.02 | 0.76*** | 0.15 | −0.22 | 0.55 | (0.02) | (0.25) | (0.31) | (0.45) | (0.36) | [0.10] | [0.18] | [0.06] | [0.40] | 0.04 |
| 1,440-min | 587 | −0.01 | 0.35 | 3.30* | 0.68 | 0.38 | (0.15) | (1.63) | (2.00) | (2.94) | (2.38) | [0.23] | [0.94] | [0.92] | [0.36] | 0.02 |
| *(c) Bull & Bear Market Algorithm of Lunde and Timmermann (2004)* | | | | | | | | | | | | | | | | |
| 5-min | 587 | 0.00 | 0.70*** | 0.29** | 0.07 | −0.02 | (0.01) | (0.09) | (0.13) | (0.17) | (0.14) | [0.01] | [0.70] | [0.00] | [0.12] | 0.10 |
| 10-min | 587 | 0.00 | 0.83*** | 0.39** | 0.11 | −0.03 | (0.01) | (0.13) | (0.18) | (0.24) | (0.20) | [0.04] | [0.65] | [0.01] | [0.12] | 0.09 |
| 15-min | 587 | −0.01 | 1.15*** | 0.66*** | 0.15 | −0.07 | (0.01) | (0.15) | (0.20) | (0.27) | (0.22) | [0.04] | [0.54] | [0.00] | [0.02] | 0.11 |
| 30-min | 587 | 0.00 | 1.57*** | 0.66** | 0.00 | −0.11 | (0.02) | (0.21) | (0.30) | (0.40) | (0.33) | [0.01] | [0.84] | [0.00] | [0.09] | 0.10 |
| 60-min | 587 | 0.03 | 0.31 | 0.65* | 0.55 | 0.02 | (0.02) | (0.24) | (0.34) | (0.46) | (0.38) | [0.40] | [0.37] | [0.65] | [0.22] | 0.02 |
| 1,440-min | 587 | −0.03 | 2.97* | −0.68 | −13.39 | 1.00 | (0.15) | (1.56) | (2.18) | (8.35) | (2.41) | [0.15] | [0.10] | [0.06] | [0.61] | 0.01 |

*Notes*: Table 4 reports estimates of Equation (3) (i.e., $r_t = \alpha + [\beta_1(D_t^m D_t^+) + \beta_2(D_t^m D_t^-) + \beta_3(D_t^b D_t^+) + \beta_4(D_t^b D_t^-)]\Delta T_t + e_t$) using robust MM-estimates. $r_t$ denotes Bitcoin returns. $\Delta T_t$ denotes the change in USD₮ supply. $\Delta T_t$ denotes the change in the USD₮ supply. $D_t^m$ and $D_t^b$ are dummy variables that are equal to one when the event date is a minting or a burning event and equal to zero otherwise. $D_t^+$ and $D_t^-$ are dummy variables that are equal to one (zero) when: the Crypto Fear & Greed Index points to extreme greed or greed (extreme fear or fear) in Panel (a); the Sentix Bitcoin Institutional Investor Sentiment Indicator surveys more bulls than bears (more bears than bulls) in Panel (b); and the Lunde and Timmermann filter (2012) indicates a bullish (bearish) market in Panel (c). The sample extends from October 6, 2014 18:54:05 UTC to January 9, 2021 13:20:09 UTC. Standard errors appear in parentheses. P-values from Wald tests (F-statistics) appear in square brackets. *, **, *** indicate significance at the 10%, 5% and 1% levels.

**Table 5: Response of Bitcoin returns to USD₮ supply change events, controlling for minting events and Whale Alerts on Twitter**

| Event Window | Obs. | α | $\beta_1$ | $\beta_2$ | $\beta_3$ | $\beta_4$ | $\alpha^{SE}$ | $\beta_1^{SE}$ | $\beta_2^{SE}$ | $\beta_3^{SE}$ | $\beta_4^{SE}$ | [$\beta_1=\beta_2$] | [$\beta_3=\beta_4$] | [$\beta_1=\beta_3$] | [$\beta_2=\beta_4$] | Adj. $R^2$ |
|---|---|---|---|---|---|---|---|---|---|---|---|---|---|---|---|---|
| *(a) Crypto Fear and Greed Index of Alternative.me* | | | | | | | | | | | | | | | | |
| 5-min | 587 | 0.01 | 0.66*** | 0.16 | 0.71* | −2.14 | (0.01) | (0.10) | (0.11) | (0.39) | (3.23) | [0.00] | [0.38] | [0.90] | [0.48] | 0.16 |
| 10-min | 587 | 0.07*** | 0.57*** | 0.02 | 0.17 | −21.96* | (0.02) | (0.14) | (0.15) | (0.53) | (11.13) | [0.00] | [0.05] | [0.45] | [0.05] | 0.10 |
| 15-min | 587 | 0.06** | 1.12*** | 0.01 | −0.43 | −25.35* | (0.02) | (0.16) | (0.18) | (0.62) | (12.91) | [0.00] | [0.05] | [0.01] | [0.05] | 0.17 |
| 30-min | 587 | 0.09** | 1.08*** | −0.14 | 1.01 | −9.12 | (0.03) | (0.26) | (0.28) | (0.97) | (8.12) | [0.00] | [0.21] | [0.94] | [0.27] | 0.10 |
| 60-min | 587 | 0.09** | 0.32 | −0.15 | 1.52 | −5.35 | (0.04) | (0.30) | (0.33) | (1.15) | (9.67) | [0.25] | [0.48] | [0.30] | [0.59] | 0.01 |
| 1,440-min | 587 | 0.28 | −1.09 | 0.37 | 10.58 | −25.48 | (0.25) | (1.87) | (2.04) | (7.09) | (59.48) | [0.56] | [0.55] | [0.10] | [0.66] | 0.00 |
| *(b) Sentix Bitcoin Institutional Investor Sentiment Indicator* | | | | | | | | | | | | | | | | |
| 5-min | 587 | 0.01 | 0.71*** | 0.13 | 0.74 | −0.22 | (0.01) | (0.10) | (0.14) | (0.49) | (0.54) | [0.00] | [0.18] | [0.96] | [0.51] | 0.18 |
| 10-min | 587 | 0.05** | 0.56*** | 0.37* | 0.51 | −0.30 | (0.02) | (0.15) | (0.19) | (0.69) | (0.75) | [0.38] | [0.42] | [0.94] | [0.37] | 0.09 |
| 15-min | 587 | 0.04* | 0.92*** | 0.24 | −0.47 | −0.34 | (0.02) | (0.18) | (0.23) | (0.84) | (0.91) | [0.01] | [0.91] | [0.10] | [0.53] | 0.12 |
| 30-min | 587 | 0.08** | 0.99*** | 0.19 | 1.07 | 1.17 | (0.04) | (0.27) | (0.35) | (1.28) | (1.38) | [0.05] | [0.96] | [0.95] | [0.49] | 0.07 |
| 60-min | 587 | 0.11 | 0.42 | −0.29 | 2.35 | −2.70* | (0.04) | (0.31) | (0.40) | (1.45) | (1.57) | [0.12] | [0.02] | [0.19] | [0.13] | 0.06 |
| 1,440-min | 587 | 0.23 | −0.63 | 1.30 | 34.44* | −1.58 | (0.27) | (1.92) | (2.50) | (20.10) | (9.85) | [0.49] | [0.01] | [0.00] | [0.77] | 0.05 |
| *(c) Bull & Bear Market Algorithm of Lunde and Timmermann (2004)* | | | | | | | | | | | | | | | | |
| 5-min | 587 | 0.00 | 0.67*** | 0.29** | 0.56 | −1.84 | (0.01) | (0.10) | (0.14) | (0.37) | (3.45) | [0.02] | [0.49] | [0.77] | [0.54] | 0.17 |
| 10-min | 587 | 0.04** | 0.64*** | 0.35* | 0.26 | −4.32 | (0.02) | (0.15) | (0.19) | (0.51) | (4.81) | [0.20] | [0.34] | [0.45] | [0.33] | 0.10 |
| 15-min | 587 | 0.01 | 1.21*** | 0.60** | −0.26 | 7.56 | (0.02) | (0.18) | (0.23) | (0.62) | (5.81) | [0.02] | [0.18] | [0.02] | [0.23] | 0.20 |
| 30-min | 587 | 0.03 | 1.32*** | 0.55 | 1.32 | 9.92 | (0.04) | (0.27) | (0.35) | (0.94) | (8.84) | [0.06] | [0.33] | [1.00] | [0.29] | 0.10 |
| 60-min | 587 | 0.09** | −0.03 | 0.37 | 0.75 | 10.95 | (0.04) | (0.31) | (0.41) | (1.11) | (10.36) | [0.40] | [0.33] | [0.48] | [0.31] | 0.02 |
| 1,440-min | 587 | 0.16 | 1.27 | −1.97 | 15.41** | −49.08 | (0.27) | (1.93) | (2.56) | (6.81) | (63.83) | [0.27] | [0.31] | [0.04] | [0.46] | 0.01 |

Notes: Table 5 reports estimates of Equation (4) (i.e., $r_t = \alpha + [(\beta_1 D_t^+ + \beta_2 D_t^-)D_t^{wa} + (\beta_3 D_t^+ + \beta_4 D_t^-)(1 - D_t^{wa})]\Delta T_t D_t^m + e_t$) using robust MM-estimates. $r_t$ denotes Bitcoin returns. $\Delta T_t$ denotes the change in the USD₮ supply. $D_t^m$ is a dummy variable that is equal to one when the event date is a minting or burning event and zero otherwise. $D_t^+$ and $D_t^-$ are dummy variables that are equal to one (zero) when: the Crypto Fear & Greed Index points to extreme greed or greed (extreme fear or fear) in Panel (a); the Sentix Bitcoin Institutional Investor Sentiment Indicator surveys more bulls than bears (more bears than bulls) in Panel (b); and the Lunde and Timmermann filter (2012) indicates a bullish (bearish) market in Panel (c). $D_t^{wa}$ is a dummy variable that is equal to one when Ethereum and Tron USD₮ minting events are accompanied by a Whale Alert tweet and equal to zero otherwise. The sample extends from October 6, 2014 18:54:05 UTC to January 9, 2021 13:20:09 UTC. Standard errors appear in parentheses. P-values from Wald tests (F-statistics) appear in square brackets. *, **, *** indicate significance at the 10%, 5% and 1% levels.